\documentclass[prd,floatfix,preprintnumbers,letterpaper]{revtex4}
\usepackage{graphicx}
\usepackage{amsmath,amssymb}
\input{epsf}
\usepackage{epsf}
\usepackage{color}

\newcommand {\ga} {\ {\raise-.5ex\hbox{$\buildrel>\over\sim$}}\ }
\newcommand {\la} {\ {\raise-.5ex\hbox{$\buildrel<\over\sim$}}\ } 
\begin{document}

\title{Direct cosmographic reconstruction of the quintessence potential}

\author{Saikat Chakraborty}
\affiliation{Center for Space Research, North-West University, Potchefstroom 2520, South Africa}
\email{saikat.chakraborty@nwu.ac.za}

\author{Peter K.S. Dunsby}
\affiliation{Department of Mathematics and Applied Mathematics, University of Cape Town,
Rondebosch 7701, Cape Town, South Africa}
\email{peter.dunsby@uct.ac.za}

\author{Robert J. Scherrer}
\affiliation{Department of Physics and Astronomy, Vanderbilt University,
Nashville, TN 37235}
\email{robert.scherrer@vanderbilt.edu}

\begin{abstract}
We derive expressions for the first
and second derivatives of the
quintessence potential $V(\phi)$, in terms
of $\lambda = -V^{\prime}/V$ and $\Gamma = (V^{\prime \prime}/V)/(V^\prime/V)^2$,
as functions
of the quintessence density fraction $\Omega_\phi$ and the cosmographic parameters $q$, $j$, and $s$.  Our mapping is not explicitly a function
of the equation of state parameter $w$.
We use these results, along with recent observational data, to derive expansions of $V(\phi)$ about the present-day value of the scalar field, $\phi_0$.
\end{abstract}

\maketitle
\section{Introduction}
The origin of the late-time accelerated expansion of the universe remains an open problem in cosmology \cite{Riess,Perlmutter}. Within the standard $\Lambda$CDM framework, this acceleration is attributed to a cosmological constant, $\Lambda$, corresponding to a constant vacuum energy density with the equation of state $w=-1$. Despite its phenomenological success, $\Lambda$CDM faces well-known theoretical challenges, and current observations do not exclude the possibility of dynamical dark energy. In particular, recent results from the Dark Energy Spectroscopic Instrument (DESI) have suggested potential deviations from a pure cosmological constant, motivating further investigation of alternative models \cite{DESIDR1,DESIDR2,DESIDR2p2}.

The dark energy component can be characterized by its equation of state parameter $w$, which is defined as the ratio of the 
dark energy pressure to its density $w=p/\rho$. The special case $w=-1$ and constant $\rho$, corresponds to a cosmological constant.

One set of alternatives to $\Lambda$ consists of quintessence models, in which the dark energy originates from a scalar field $\phi$ with an associated potential $V(\phi)$ \cite{RatraPeebles,Wetterich,Ferreira,CLW,Ferreira2,CaldwellDaveSteinhardt,Liddle,SteinhardtWangZlatev}.
(See Ref. \cite{Copeland1} for a review). In general $w$ evolves with time in such models. Although $w$ arises naturally in the analysis of scalar field dark energy, it suffers from several drawbacks. It is not related to a directly-observable quantity. Furthermore, it has long been known that even perfect knowledge of supernova luminosity distances is insufficient to determine the exact evolution of $w$ \cite{Maor1,Wolf1}. This limitation motivates the consideration of alternative, more directly observable parameterizations of the expansion history. 

An alternative is to parametrize the expansion history directly in terms of the so called cosmographic parameters, which correspond to derivatives of the scale factor $a$ \cite{Sahni}. Cosmography provides a model-independent framework in which the expansion of the universe is characterized by kinematic quantities derived from the scale factor, including the Hubble parameter $H\equiv\dot{a}/a$, the deceleration parameter 
$q\equiv - \ddot{a}/a H^2$ and higher-order derivatives such as the jerk $j\equiv\dddot{a}/a H^3$ and snap $s\equiv \ddddot{a}/aH^4$. These quantities can, in principle, be inferred directly from observations without assuming a specific dark energy model. This series can be continued to arbitrarily order in the time derivative, but in practice these higher-order expressions are inaccessible to current observations (see Ref. \cite{Dunsby} for a review).

Some previous work has focused on relating statefinder parameters to quintessence dynamics. Nair, Jhingan, and Jain \cite{Nair} used Gaussian processes to derive the luminosity distance as a function of redshift from cosmographic variables and then worked backwards to derive the scalar field kinetic energy and potential as a function of redshift. Dubey et al. \cite{Dubey} derive scalar field dynamics using a parametrization of the deceleration parameter.  Other approaches to reconstruction of the quintessence potential can be found in Refs. \cite{Gerke,Sahlen,Mukherjee,Elizalde}.

In this work, we adopt an alternative approach to derive an exact relation between the quintessence potential and the statefinder parameters, thereby avoiding any explicit dependence on $w$.

In what follows, we obtain expressions for  $\lambda = -V^{\prime}/V$ and $\Gamma = (V^{\prime \prime}/V)/(V^\prime/V)^2$, in terms of the quintessence density fraction $\Omega_\phi$ and the cosmographic parameters. In Sec. IV, we employ recent observational data sets, together with these expressions for the derivatives of the potential, to reconstruct $V(\phi)$ near the present-day value of $\phi$.  In Sec. V, we discuss the sensitivity of our estimates to the errors in the cosmographic parameters.  Our conclusions are summarized in Sec. VI.
\section{Quintessence model}
We will assume that the dark energy is provided by a minimally-coupled scalar field, $\phi$ and associated potential $V(\phi)$. The pressure and energy density of the scalar field are given by the standard expressions
\begin{equation}
p = \frac{\dot \phi^2}{2} - V(\phi)\;,
\end{equation}
and
\begin{equation}
\label{rhodense}
\rho = \frac{\dot \phi^2}{2} + V(\phi)\;,
\end{equation}
with equation of motion (which follows from energy conservation) given by
\begin{equation}
\label{motionq}
\ddot{\phi}+ 3H\dot{\phi} + \frac{dV}{d\phi} =0\;,
\end{equation}
where the Hubble parameter $H$ is
\begin{equation}
\label{H}
H = \sqrt{\rho/3},
\end{equation}
and $\rho$ is the total density in units for which $8 \pi G = 1$.

These equations can be used to derive a simple relation between the potential and its derivative
$V^\prime \equiv dV/d\phi$, $w$, and $\Omega_\phi$, namely \cite{zlatev}
\begin{equation}
\label{motion1}
-\frac{V^\prime}{V} = \sqrt{\frac{3(1+w)}{\Omega_\phi}}\left[
1 + \frac{w^\prime}{3(1-w)(1+w)}\right],
\end{equation}
where $w^\prime$ is the derivative of $w$ with respect to $\ln(a)$.

The cosmological field equations of the quintessence model can be conveniently expressed as the following autonomous system \cite{ScherrerSen}
\begin{subequations}\label{autonomous}
\begin{align}
\frac{dw}{d\ln a} &= \lambda(1-w)\sqrt{3(1+w)\Omega_\phi}\,-3\left(1-w^2\right)\,,
\\
\frac{d\Omega_{\phi}}{d\ln a} &= -3w\Omega_\phi(1-\Omega_\phi)\,,\label{Omega'-eq}
\\
\frac{d\lambda}{d\ln a} &= -\sqrt{3}(\Gamma-1)\lambda^2\,\sqrt{(1+w)\Omega_\phi}\,,\label{lambda'-eq}
\end{align}
\end{subequations}
with $\lambda\equiv-\frac{V'}{V},\,\Gamma\equiv\frac{V''/V}{(V'/V)^2}$.

Furthermore, the relation between the cosmographic deceleration parameter $q$, jerk parameter $j$, and the quintessence parameters is
given by
\begin{eqnarray}
\label{q}
q &=& \frac{1}{2} + \frac{3}{2} w \Omega_\phi,\\
\label{j}
j &=& 1 + \frac{3}{2} \Omega_\phi(3w +3w^2 - w^\prime).
\end{eqnarray}
The cosmographic parameters themselves follow a kind of hierarchy
\begin{subequations}\label{CP_rel}
    \begin{eqnarray}
        j &=& 2q^{2} + q - \frac{dq}{d\ln a}\,,\label{q'-eq}
        \\
        s &=& \frac{dj}{d\ln a} - j(2 + 3q)\,,\label{j'-eq}
    \end{eqnarray}
\end{subequations}
where $s$ is the next cosmography parameter up the cosmographic ladder, the snap parameter.

Our objective is to express the potential slope parameters, $\lambda$ and $\Gamma$, in terms of cosmographic variables—which can, in principle, be measured directly—and $\Omega_\phi$, which is relatively well constrained. In doing so, we aim to eliminate $w$ and $w'$, as these quantities can only be inferred indirectly. By using Eq. (\ref{j}), we can eliminate $w'$ from Eq. (\ref{motion1}).

to obtain
\begin{equation}
\label{motion2}
\lambda\equiv-\frac{V^\prime}{V} = \sqrt{\frac{3(1+w)}{\Omega_\phi}}\left[
\frac{1}{1-w} - \frac{2}{9} \frac{j-1}{\Omega_\phi(1-w)(1+w)}\right],
\end{equation}
and then Eq. (\ref{q}) can be used to eliminate $w$, giving our final result:
\begin{equation}
\label{motion3}
\lambda = \frac{9\Omega_{\phi} + 6q - 2j - 1}{(3\Omega_{\phi}-2q+1)\sqrt{3\Omega_{\phi}+2q-1}}\,.
\end{equation}
We can go a step further by expressing the higher-order potential slope parameter, $\Gamma$, in terms of the cosmographic parameters and $\Omega_\phi$. To do this, we differentiate Eq.~\eqref{motion3} and use Eqs.~\eqref{Omega'-eq} and \eqref{lambda'-eq} to substitute for $\frac{d\Omega}{d\ln a}$ and $\frac{d\lambda}{d\ln a}$. We also replace $\frac{dq}{d\ln a}$ and $\frac{dj}{d\ln a}$ using Eqs.~\eqref{q'-eq} and \eqref{j'-eq}. The resulting expression is
\begin{widetext}
\begin{equation}\label{motion4}
    \Gamma = \frac{(3\Omega_\phi-2q+1) \left[4j^2+8j(2q+3\Omega_\phi-2)+(2q+3\Omega_\phi)(6q(2q+3\Omega_\phi+5)+4s+45\Omega_\phi)-66q-4s-90\Omega_\phi+19\right]}{2(3\Omega_\phi+2q-1)(9\Omega_{\phi} + 6q - 2j - 1)^2}\,.
\end{equation}
\end{widetext}

The expressions \eqref{motion3} and \eqref{motion4} are exact and hold for all quintessence potentials. In principle, one could further reconstruct all the higher-order potential slope parameters in terms of higher-order cosmographic parameters.

\section{$j=1$ does \emph{not} necessarily imply a flat potential}\label{sec:j=1}

In the statefinder diagnostic \cite{Sahni,Nair,Dubey}, the cosmographic condition $j=1$ is commonly taken to characterize $\Lambda$CDM evolution. However, setting $j=1$ in Eq.~\eqref{motion3} yields
\begin{equation}\label{motion_LCDM_1}
\frac{V'}{V} = - 3,\frac{\sqrt{3\Omega_{\phi}+2q-1}}{(3\Omega_{\phi}-2q+1)} ,.
\end{equation}
A specific case arises when $\Omega_\phi$ evolves as $\Omega_\phi=\frac{1}{3}(1-2q)$, for which the potential becomes flat, $V'(\phi)=0$. This corresponds to the $\Lambda$CDM model. From Eq.~\eqref{q}, it follows that this condition implies $w=-1$. In this limit, the higher-order potential slope parameter $\Gamma$, as defined below Eq.~\eqref{autonomous}, is not well defined.
More generally, however, the condition $j=1$ does not guarantee a flat potential. As emphasized by Chakraborty et al.~\cite{Chakrabortyetal}, a model with $j=1$ does not necessarily imply $w=-1$; rather, it corresponds to the behavior
\begin{equation}
w = - \frac{a^3}{a^3 + C},
\end{equation}
where $C$ is a constant.
This result has a simple physical interpretation.  An equation of state parameter of this form corresponds to a density that scales as the sum of a matter-like component ($\rho \propto a^{-3}$) and a constant-density component,
and the matter-like component can be absorbed into a redefinition of $\Omega_M$.

Thus, this family of models is indistinguishable from the $\Lambda$CDM model at the background level, as far as cosmographic data are concerned. However, these models are not identical to $\Lambda$CDM at the perturbative level, as quintessence does not cluster gravitationally. Thus, a model of this kind will consist of a cosmological constant, standard cold dark matter, and a non-clustering component with a density scaling as $a^{-3}$, 
resulting in a suppression of perturbation growth that can be distinguished observationally from $\Lambda$CDM.

Evolution of this sort is easily achieved in the context of quintessence.  It has long been known that an exponential potential of the form
\begin{equation}
\label{Vexp}
V(\phi) = V_1 \exp(-\lambda \phi),
\end{equation}
can lead to attractor scaling behavior, in which $\phi$ evolves in such a way that $w$ approaches the value of the equation of state parameter for the dominant background fluid, $w_B$ \cite{CLW,Ferreira2}.  The condition for this scaling behavior is $\lambda^2 > 3(1+w_B)$, in which case
the attractor solution for $\Omega_\phi$ is
$\Omega_\phi = 3(1+w_B)/\lambda^2$.  When cold dark matter
is dominant, a potential of the form given by Eq. (\ref{Vexp})
yields a quintessence fluid with $w = 0$ and $j=1$.

We can
then add a second (constant) component to the potential, so
that
\begin{equation}
\label{Vexp}
V(\phi) = V_0 + V_1 \exp(-\lambda \phi).
\end{equation}
When combined with a cold dark matter background, this leads to the behavior described above: $j=1$, with the quintessence field effectively behaving as a sum of a (non-clustering) matter component and a vacuum energy density. Models of this type have been proposed as candidates for dark energy in Ref.~\cite{ChangScherrer}.
However, strictly speaking, the potential given by Eq.~\eqref{Vexp} is not the most general form capable of producing this behavior. More generally, one can also include an additional $e^{\lambda\phi}$ term, leading to a potential of the form
\begin{equation}
\label{Vcosh}
V(\phi) = V_0 + V_1 \cosh(\lambda\phi)\,.
\end{equation}
Such a model was introduced in \cite{BruniCosh} as an effective phenomenological model for the unified dark sector in the quintessence framework. In the limit of $\lambda\phi\to-\infty$, the potential \eqref{Vcosh} essentially reduces to \eqref{Vexp}. 

\section{A direct cosmographic reconstruction of the quintessence potential}\label{sec:reconstruction}

The cosmographic reconstruction presented here is both distinct from and more direct than standard approaches applied to quintessence potentials. In conventional cosmographic reconstruction, one computes successive derivatives of the potential $V(\phi)$ with respect to redshift $z$ and expresses them in terms of cosmographic parameters \cite{DunsbyLuongo}. However, reconstructing the potential $V(\phi)$ itself then additionally requires knowledge of the field evolution, $\phi = \phi(z)$.
In contrast, our approach directly evaluates derivatives of the potential with respect to the scalar field. This allows for a more straightforward reconstruction of the potential around a given field value $\phi_0$ using a Taylor expansion
\begin{align}\label{potential_taylor}
   V(\phi) &= V(\phi_0) + V'(\phi_0)(\phi-\phi_0) + \frac{1}{2}V''(\phi_0)(\phi-\phi_0)^2 + ..... \nonumber\\
    &= V(\phi_0)\left[1-\lambda_0(\phi-\phi_0)+\frac{1}{2}\lambda_0^2\,\Gamma_0(\phi-\phi_0)^2+.....\right]\,. 
\end{align}

For our purpose, we consider two cosmographic studies: one after DESI DR1 \cite{cosmography_DESI1} and another after DESI DR2 \cite{cosmography_DESI2}. We take the best fit values of $q_0,j_0,s_0$ from \cite[Table 2]{cosmography_DESI1} and \cite[Table 1]{cosmography_DESI2}, respectively. To calculate $\lambda_0,\Gamma_0$ we still need $\Omega_{\phi0}$, which is $\Omega_{\rm DE0}$ in our case. Model-independent cosmographic studies do not measure $\Omega_{m0}$. However, many other observations constrain $\Omega_{m0}$ to be $\approx0.3$, which gives $\Omega_{\rm DE 0}=\Omega_{\phi0}=0.7$. We take this value of $\Omega_{\phi0}$ for our subsequent analysis.

Table \ref{tab:cosm_values_dr1} and Table \ref{tab:cosm_values_dr2} list the best fit values of $q_0,j_0,s_0$ as can be found in \cite[Table 2]{cosmography_DESI1} and \cite[Table 1]{cosmography_DESI2} respectively, along with the respective values of $\lambda_0$ and $\Gamma_0$ that can be calculated for the fiducial value $\Omega_{\phi0}=0.7$.
\begin{table}[h]
    \centering
    \begin{tabular}{|c|c|c|c|c|c|}
    \hline 
        Data & $q_0$ & $j_0$ & $s_0$ & $\lambda_0$ & $\Gamma_0$
        \\
        \hline 
        DESY5 & $-0.503$ & $0.97$ & $-0.56$ & $0.272$ & $11.385$
        \\
        \hline 
        Pantheon + & $-0.465$ & $0.85$ & $-0.33$ & $0.487$ & $7.694$
        \\
        \hline
    \end{tabular}
    \caption{Cosmographic values taken from \cite[Table 2]{cosmography_DESI1}. $\Omega_{\phi0}=0.7$ is assumed.}
    \label{tab:cosm_values_dr1}
\end{table}

\begin{table}[h]
    \centering
    \begin{tabular}{|c|c|c|c|c|c|}
    \hline 
        Method & $q_0$ & $j_0$ & $s_0$ & $\lambda_0$ & $\Gamma_0$
        \\
        \hline 
        Taylor ($z\leq1$) & $-0.41$ & $1.48$ & $1.64$ & $-0.058$ & $2760.69$  
        \\
        \hline 
        Pad\'e$_{(2,2)}$ & $-0.397$ & $0.73$ & $0.55$ & $0.677$ & $7.410$
        \\
        \hline 
        Chebyshev & $-0.472$ & $0.59$ & $-0.48$ & $0.806$ & $4.076$
        \\
        \hline
    \end{tabular}
    \caption{Cosmographic values taken from \cite[Table 1]{cosmography_DESI2}. $\Omega_{\phi0}=0.7$ is assumed.}
    \label{tab:cosm_values_dr2}
\end{table}

\begin{figure}[h]
    \centering
    \includegraphics[width=0.7\linewidth]{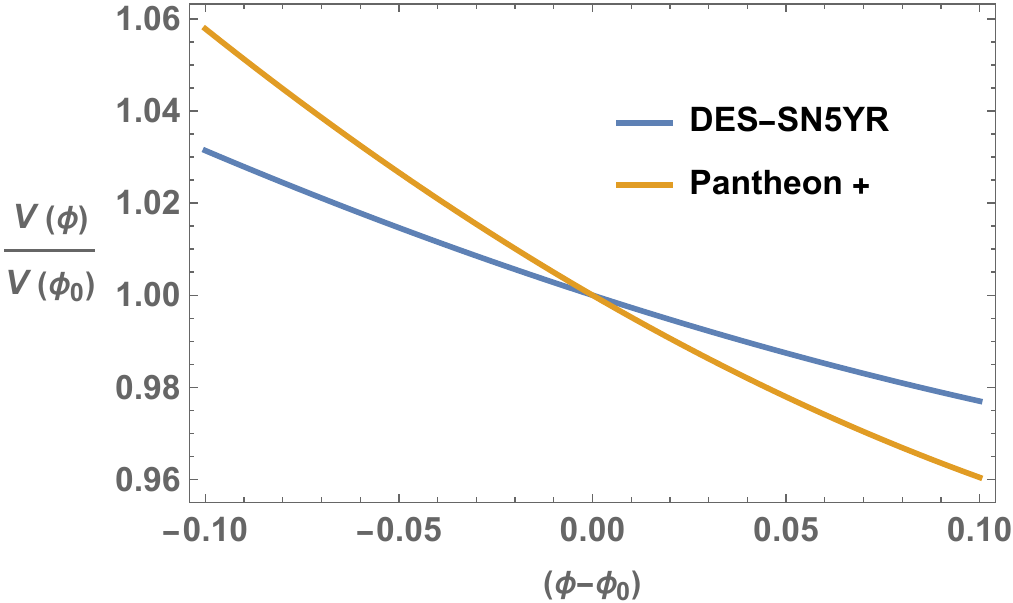}
    \caption{Quintessence potential reconstruction with DESI DR1 data}
    \label{fig:potentials_dr1}
\end{figure}

Based on the values of $\lambda_0,\Gamma_0$ appearing in Table \ref{tab:cosm_values_dr1}, we can reconstruct the quintessence potential as
\begin{widetext}
\begin{align}
V(\phi) &\approx V(\phi_0)\left[1-0.272(\phi-\phi_0)+0.420(\phi-\phi_0)^2+....\right] \qquad (\text{DESY5})\,,
\\
V(\phi) &\approx V(\phi_0)\left[1-0.487(\phi-\phi_0)+0.914(\phi-\phi_0)^2+....\right] \qquad (\text{Pantheon +})\,.
\end{align}
\end{widetext}
The corresponding potentials are shown in Fig.\ref{fig:potentials_dr1}.

\begin{figure}[h]
    \centering
    \includegraphics[width=0.7\linewidth]{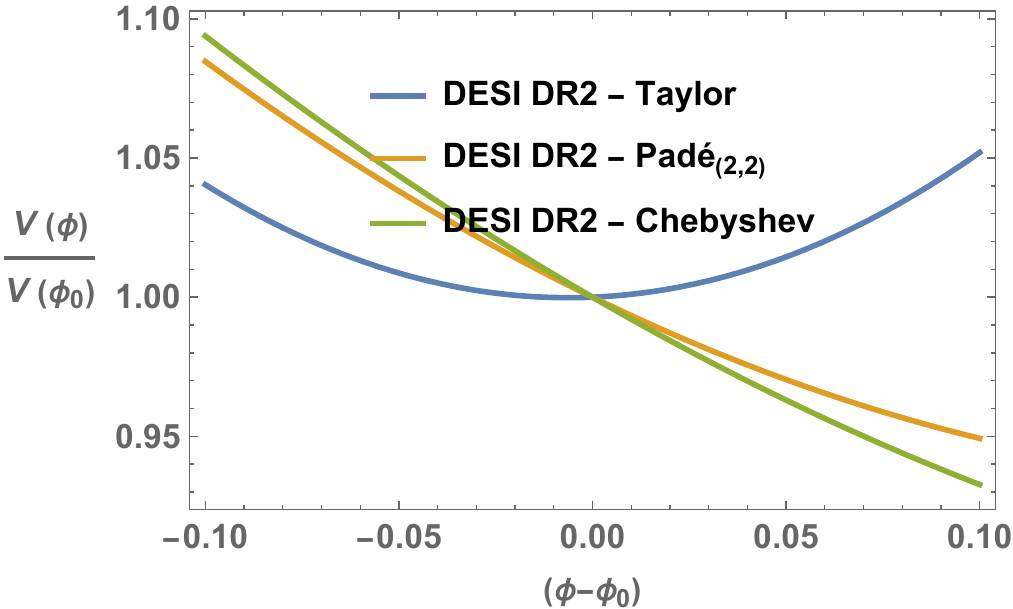}
    \caption{Quintessence potential reconstruction with DESI DR2 data}
    \label{fig:potentials_dr2}
\end{figure}

On the other hand, based on the values of $\lambda_0,\Gamma_0$ appearing in Table \ref{tab:cosm_values_dr2}, we can reconstruct the quintessence potential as
\begin{widetext}
\begin{align}
V(\phi) &\approx V(\phi_0)\left[1+0.058(\phi-\phi_0)+4.620(\phi-\phi_0)^2+....\right]\,, \qquad (\text{DESI DR2 - Taylor})\,,
\\
V(\phi) &\approx V(\phi_0)\left[1-0.677(\phi-\phi_0)+1.697(\phi-\phi_0)^2+....\right]\,, \qquad (\text{DESI DR2 - Pad\'e$_{(2,2)}$})\,,
\\
V(\phi) &\approx V(\phi_0)\left[1-0.806(\phi-\phi_0)+1.325(\phi-\phi_0)^2+....\right]\,, \qquad (\text{DESI DR2 - Chebyshev})\,.
\end{align}
\end{widetext}
The corresponding potentials are shown in Fig.\ref{fig:potentials_dr2}.

\section{Sensitivity to the cosmographic parameters} \label{sec:sensitivity}

Of the cosmographic parameters considered here, the best-measured is $q$, followed by $j$ and then $s$, while $\Omega_\phi$ is better constrained than any of these.  Consider, then, the sensitivity of the slope of potential, $\lambda$, to the error in $j$.  Taking $j = j_0 + \Delta j$ in Eq. (\ref{motion3}), we obtain the relative error in $\lambda$:
\begin{equation}
\frac{\Delta \lambda}{\lambda} = - \frac{2 \Delta j}{9 \Omega_\phi + 6 q - 2j_0 -1}.
\end{equation}
It is informative to rewrite this expression in terms of the corresponding value for $w$:
\begin{equation}
\frac{\Delta \lambda}{\lambda} = \frac{\Delta j}{(j_0-1)  + \frac{9}{2}
\Omega_\phi(1+w)}.
\end{equation}
Clearly, this error is largest when the data correspond to a nearly $\Lambda$CDM model with $j_0 = 1$ and $w = -1$, giving
$\lambda$ near zero.  Conversely, the further the data deviate from pure $\Lambda$CDM, the less sensitive the value of $\lambda$ will be to the exact value of $j$.

\section{Conclusions}

We have demonstrated that the cosmographic parameters, when combined with $\Omega_\phi$, can be mapped directly onto the first and second derivatives of the quintessence potential, expressed through $\lambda = -V'/V$ and $\Gamma = (V''/V)/(V'/V)^2$. While Eqs.~(\ref{motion3}) and (\ref{motion4}) are valid at arbitrary redshift, their dependence on $\Omega_\phi$ makes them most useful at the present epoch, where observational constraints on $\Omega_\phi$ are strongest.
By combining these expressions, cosmographic parameters inferred from recent observations can be used to construct a Taylor expansion of $V(\phi)$ about the present-day field value $\phi_0$. As expected, the resulting reconstructions favor relatively flat potentials. It is important to emphasize that the determination of $\lambda$ is more robust than that of $\Gamma$: the former depends only on $q$ and $j$, which are relatively well constrained, whereas the latter requires knowledge of $s$, which is less well determined. This difference is reflected in Sec.~IV, where $\lambda_0$ is consistently constrained, while $\Gamma_0$ exhibits significantly larger uncertainty.
The primary advantage of our approach is that it avoids direct dependence on the equation-of-state parameter $w$, which is traditionally used in dark energy studies. Since the scale factor depends on an integral over $w$, this parameter is not ideally suited for observational reconstruction. In this sense, our direct mapping from cosmographic parameters to quintessence parameters effectively bypasses this intermediate step and provides a more direct framework for analyzing dark energy.

\acknowledgments

S.C. acknowledges funding support from the Second Century Fund (C2F), Chulalongkorn University, Thailand. P.D. acknowledges First Rand Bank (SA) and the University of Cape Town Research Committee for financial support.

\end{document}